\documentclass{ws-procs9x6}

\newcommand{\ltsim}{\mbox{{\raisebox{-0.4ex}{$\stackrel{<}{{\scriptstyle\sim}}
$}}}}

\begin{document}

\title{A large area search for radio-loud quasars within the epoch of
reionization}

\author{M.J.~Jarvis$^{1}$, S.~Rawlings$^{1}$, F.E.~Barrio$^{1}$,\\
  G.J.~Hill$^{2}$, A.~Bauer$^{2}$, S.~Croft$^{3}$}

\address{$^{1}$Astrophysics, Department of Physics, Keble Road, Oxford, OX1
  3RH, UK.
$^{2}$McDonald Observatory, University of Texas at Austin, 1
  University Station C1402, Austin, TX 78712-0259, USA.$^{3}$IGPP, Lawrence Livermore National Laboratory, L-413, 7000 East
  Ave., Livermore, CA 94550, USA\\
E-mail: mjj@astro.ox.ac.uk}


\maketitle

\abstracts{
The Universe became fully reionized, and observable optically,
at a time corresponding to redshift $z \sim 6.5$,
so it is only by studying the HI and molecular absorption lines
against higher-redshift, radio-loud sources that one can hope to make detailed
studies of the earliest stages of galaxy formation. At present
no targets for such studies are known. In these proceedings we
describe a survey which is underway to find radio-loud quasars at $z >
6.5$, and present broad-band SEDs of our most promising candidates. 
}

\section{Introduction}
The epoch of reionization has now been
discovered as a protracted period reaching from $z \sim 20 \rightarrow
6.5$ \cite{kog,beck}.
However, prior to $z \sim 6.5$ galaxy formation was already well underway
(e.g. [3]).
It is essentially impossible to study this `grey age' at optical
wavelengths, but great progress can be expected
if radio and millimetre telescopes can be targeted on quasars
observed within the reionization epoch. 

Radio-loud targets allow absorption studies that can probe the
evolving neutral and molecular content of the high-$z$ Universe \cite{carilli}, and radio HI absorption is the
{\it only} way of probing the neutral gas which goes on to form
stars. 
We could begin these studies with current facilities (e.g.\ the
GBT and GMRT), and with the next generation of large radio telescopes, such
as the LOFAR and the SKA, we will easily be able reach depths of lower
luminosity radio sources and still detect 21~cm absorption.
Unfortunately, there are currently no known $z > 6.5$
radio-loud objects.

This is because such objects are rare, $\ll 1$ per cent of the radio population.
Interest in pursuing them was
dampened by the claim of a much sharper cut-off in their
redshift distribution\cite{shaver} than earlier work\cite{dp90} had suggested. Jarvis \& Rawlings\cite{jr00} and Jarvis et al.\cite{Jea01}
have re-examined all the evidence concerning this
redshift cutoff,
obtaining results strongly favouring a fairly gradual decline with redshift.

\section{Design of the survey}

Jarvis \& Rawlings emphasized the
care needed in sample selection and analysis.
Therefore, the survey is selected at low frequency to avoid losing the
highest-$z$ quasars, because of the steepening of radio spectra at high
rest-frame frequencies.  The only low-frequency (325 MHz) survey with
the required depth and sky coverage is WENSS/WISH\cite{reng,debreuck}.

The sky-area and further information about the survey can be found in
Jarvis et al.\cite{Jea04}. To summarize, there are approximately 10000
sources over an area of $\sim 1$ quarter of the sky.

\section{Eliminating low-redshift radio sources}

There is a challenging, but tractable, sifting problem to eliminate both galaxies and quasars at $z < 6$. 
This is done in four steps.

(i) We have cross-correlated the radio sample with publicly available
all-sky optical and near-IR imaging, i.e.\ SDSS, POSS and 2MASS,
along with more general searches of known objects via the literature
and NED.
From this investigation we have optical IDs for
about 67\% of the objects in the northern sample (comprising quasars, low-redshift galaxies, BL Lac objects etc.). The remaining objects have no detectable optical emission, typically
to $R \sim 21.5$. 

(ii) We have
initiated deeper targeted observations in R-band of the remaining
sources in the northern hemisphere, with IGI at the 2.7~m at McDonald Observatory. Observations down to $R \sim 23-23.5$ (depending on conditions) have shown
that we again cut the source list down by approximately 65 per cent.

These sources, generally extended objects which are
presumably $z\, \ltsim\, 2$ radio galaxies, are obviously not at $z >
6.5$, because like the $z \sim 6.5$ quasars already
known\cite{fan03}, these must have zero flux
below the redshifted Lyman limit due to the Gunn-Peterson trough.

(iii)  We use good-seeing near-IR imaging to find all the remaining
quasars, and eliminate all the remaining galaxies. 
This part of the survey has been underway since August 2003 for the
northern hemisphere, with a large allocation of time on the United
Kingdom Infrared Telescope (UKIRT). This is now 75 per cent complete with the
other 25 per cent set to be completed by August 2004.

(iv) We take Z-band or near-IR spectra and find the $z > 6.5$ quasars. UKIRT-UIST has recently showed its capability in detecting
quasar broad-emission lines in the highest redshift quasar known to date\cite{wmj03} allowing an estimate of the
black-hole mass in this quasar via the broad MgII emission line. Lyman-$\alpha$ is more
than twice as bright as MgII in the composite SDSS quasar spectrum and the huge drop blueward of Lyman-$\alpha$ due to absorption by neutral hydrogen is also a very strong signature.
Therefore, identification would be relatively easy in $\sim 20$~min 
exposures on the Hobby-Eberly telescope with its planned J-band
extension to its low-resolution spectrograph. We have also been
granted time on the Gemini-North telescope to do Z-band spectroscopy
of our highest priority candidates.

The southern survey will commence in April 2004 using the ESO
telescopes in Chile.

\section{$z > 6$ radio-loud quasar candidates}

It is already clear that our best candidates come in two flavours (Fig
1.). Flavour-1 are `textbook candidates' (top panel) with smooth JHK
spectral energy distributions (SEDs): although photometry gives only a
crude estimate, it seems very likely that these are quasars at
redshifts of at least 6 (to explain the sharp break between I and
J). Flavour-2 are `bumpy-SED' objects (bottom panel) for which the
only explanation we can find is that they are lightly-reddened quasars
at $z \sim 2.5$. As the errors show, we cannot, however, rule out the
possibility that these too are at $z > 6$, perhaps with some reddening,
so each of these must be followed up spectroscopically too.

To date we have six good $z > 6$ quasar candidates, follow-up
spectroscopy of these in the near future will provide the information
on their true nature, and hopefully provide us with the discovery of
the first $z > 6$ radio-loud quasars.

Discovery of such objects will lead to the first 21~cm absorption
observations within the epoch of reionization with the new generation
of radio telescopes operating at frequencies of $\nu < 300$~MHz,
e.g. the LOFAR.

\begin{figure}[ht]
\centerline{\epsfxsize=3.7in\epsfbox{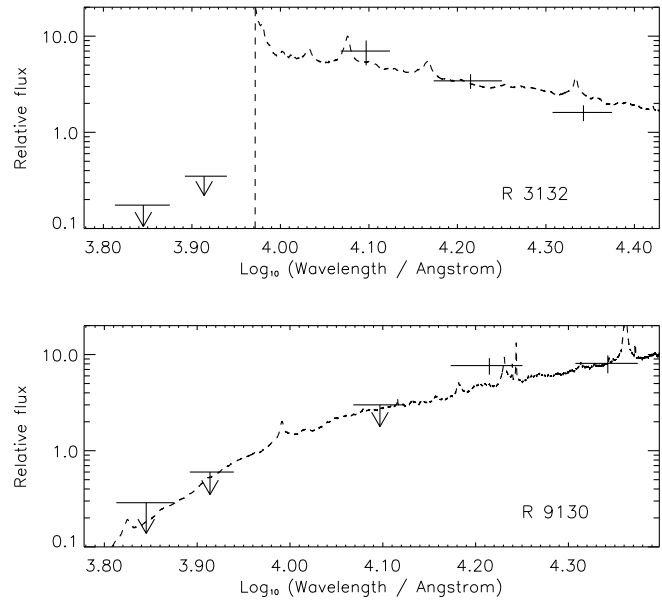}}   
\caption{SEDs for two of our best candidates, the
  dashed line is the SDSS composite quasar spectrum. (top) The quasar
  composite is redshifted to $z=6.7$, and the solid horizontal lines
  are our photometric data points from our UKIRT observations. This is
  the `textbook' SED of a $z > 6.5$ quasar. (bottom) A reddened
  (A$_{v} = 1$) quasar composite redshifted to $z=2.5$. This is our
  typical flavour-2 candidate.}
\end{figure}

%
%
%
%

\end{document}